\documentclass[a4paper,11pt]{article}
\pdfoutput=1

\usepackage{jinstpub}

\usepackage{lineno}
\usepackage{epstopdf}
\title{Instrumentation for the Study of Low Emittance Tuning and Beam Dynamics at CESR}

\author{M.G.Billing, J.A.Dobbins, M.J.Forster, D.L.Kreinick, R.E.Meller, D.P.Peterson, G.A.Ramirez,
M.C.Rendina, N.T.Rider, D.C.Sagan, J. Shanks, J.P.Sikora, M.G.Stedinger, C.R.Strohman,
H.A.Williams,\
Cornell Laboratory for Accelerator-based ScienceS and Education (CLASSE),\
Cornell University,\ 161 Synchrotron Dr., Ithaca, NY, 14850, U.S.A.\\}
\author{M.A.Palmer,\
Collider Accelerator Department, Brookhaven National Laboratory\
P.O.Box 5000, Upton, NY 11973-5000, U.S.A.\\}
\author{R.L.Holtzapple,\
Physics Department, California Polytechnic State University\
San Luis Obispo, CA 93407, U.S.A.\\}
\author{J.Flanagan,\
High Energy Accelerator Research Organization (KEK),\
1-1 Oho, Tsukuba, Ibaraki Prefecture 305-0801, Japan}

\abstract{The Cornell Electron-positron Storage Ring (CESR) has been
converted from a High Energy Physics electron-positron collider to
operate as a dedicated synchrotron light source for the Cornell High
Energy Synchrotron Source (CHESS) and to conduct accelerator physics
research as a test accelerator, capable of studying topics relevant
to future damping rings, colliders and light sources.  Some of the
specific topics that were targeted for the initial phase of
operation of the storage ring in this mode for CESR as a Test
Accelerator (CesrTA) included 1)~tuning techniques to produce low
emittance beams, 2)~the study of electron cloud development in a
storage ring and 3)~intra-beam scattering effects.  The complete
conversion of CESR to CesrTA occurred over a several year period,
described elsewhere\cite{JINST10:P07012, JINST10:P07013,
JINST11:P04025}.  In addition to instrumentation for the storage
ring, which was created for CesrTA, existing instrumentation was
modified to facilitate the entire range of investigations to support
these studies.  Procedures were developed, often requiring
coordinated measurements among different
instruments\cite{CLNS:12:2084}.  This paper describes the
instruments utilized for the study of beam dynamics during the
operation of CesrTA.  The treatment of these instruments will remain
fairly general in this paper as it focusses on an overview of the
instruments themselves.  Their interaction and inter-relationships
during sequences of observations is found in a companion paper
describing the associated measurement techniques. More detailed
descriptions and detailed operational performance for some of the
instrumentation may be found elsewhere and these will be referenced
in the related sections of this paper.}


\begin{document}
\maketitle
\flushbottom

\section{Introduction and Overview}

The initial phase of the CesrTA project was defined as having a set
of specific goals for accelerator physics research and development:

\begin{itemize}
\item The investigation of instrumentation and methodology to systematically reduce the vertical emittance of the stored beam.\cite{PRSTAB17:044003}
\item The study of the effects on the stored positron beam due to electron clouds (EC), produced by synchrotron radiation-induced photoelectrons.  The goal is to 1) characterize and quantify the production mechanisms, 2) compare these with computer simulations, 3) develop methods to mitigate the EC effects,\cite{NIMA770:141to154, PRSTAB18:041001, NIMA760:86to97} and 4) study the related beam dynamics and instabilities.\cite{IPAC14:TUPRI067}
\item The measurement and characterization of the intra-beam beam scattering effects as they lead to emittance enlargement.\cite{PRSTAB16:104401,PRSTAB17:044002,IPAC14:TUPRI035}
\end{itemize}

The diverse objectives of the CesrTA project described above require
a wide array of operating conditions, summarized in Table
\ref{tab:CESRspecs}.

\begin{table}[h]
    \centering
    \caption{\label{tab:CESRspecs} Summary of CESR operating parameters in various CesrTA configurations.}
    \vspace*{1 ex}
    \begin{tabular}{ll}
        \hline\hline
        Parameter & Specification\\
        \hline
        Circumference & 768.438~m \\
        Energy       & 1.5--5.3~GeV\\
        RF frequency & 500~MHz \\
        Typical Betatron tunes ($Q_x,Q_y$) & (14.57, 9.62) \\
        Synchrotron tune & 0.018 -- 0.072\\
        Single-bunch current & 0.1--10~mA/bunch\\
        Number of bunches & 1 -- 640 \\
        Bunch spacing & multiples of 4~ns or 14~ns\\
        Typical beam dimensions (H,W,L) & (20--100~$\mu$m, 500~$\mu$m,
        10--20~mm)\\
        \hline\hline
    \end{tabular}
\end{table}

To meet these goals a number of instruments in CESR required
significant upgrades or complete development  of new designs and
subsequent implementation.  The general description of these
instruments, which are required during coordinated measurement
sequences, are described below in the following broad categories:

\begin{itemize}
\item A major upgrade to the beam position monitor (BPM) system, which replaced an older relay-based position monitor system with individual readout modules for each BPM capable of turn-by-turn and bunch-by-bunch trajectory measurements for bunches spaced as closely as 4~nsec.
\item The installation of positron and electron vertical x-ray beam size monitors (xBSMs) designed for turn-by-turn and bunch-by-bunch beam size measurements for 4~nsec spaced bunches.
\item Implementation of a visible-light beam size monitor (vBSM) to measure the horizontal beam size for either positron or electrons, including the addition of optical elements to allow streak camera measurements of either positron or electron bunches.
\item Development of software to extract bunch-by-bunch tunes utilizing the new modules for the beam position monitors and a second method, which employed video gating of signals from a few beam position monitors from the older relay system.
\item An upgrade for the tune tracker, a device containing a feedback loop to phase lock the betatron tunes of a bunch, using either a shaker magnet or stripline kicker.  This device allows the measurement of the betatron phase advance and the horizontal-to-vertical coupling of CESR permitting their correction.
\item Installation of a new beam-stabilizing feedback system, which damps 4~nsec-spaced bunches for horizontal, vertical and longitudinal motion.
\end{itemize}

\noindent In addition these instruments utilize existing pulsed magnets or stripline kickers to excite dipole motion of the beam.  Each of these systems will be discussed in the following sections.

Note that this paper is a companion to the set of papers describing
the general modification of CESR's infrastructure to create the test
accelerator CesrTA\ \cite{JINST10:P07012, JINST10:P07013,
JINST11:P04025}. This paper is the concluding document in
preliminary conference papers describing the instrumentation,
required for the CesrTA\ Project in publications from
workshops\cite{ECLOUD10:PST07} and accelerator
conferences\cite{IPAC10:MOPE091, PAC11:WEP194, IPAC11:MOPS084} and
within the CesrTA~Phase~1 Report\cite{CLNS:12:2084}.





section
\section{Beam Position Monitors}

An upgraded beam position monitor system that provides high resolution measurement capability has been designed and deployed.  This system is capable of turn-by-turn measurements of individual bunches within bunch trains with spacings that are multiples of either 4~nsec or 14~nsec.   The system provides the ability to make closed orbit, betatron phase, coupling and dispersion measurements via synchronous detection of a driven beam.\\

\subsection{System Requirements}
The primary operational requirements for the new CESR BPM (CBPM)
system are summarizd in Table ~\ref{tab:BPMRequirements}, and
include:
\begin{itemize}
\item The ability to operate with counter-rotating beams of electrons and positrons in a single vacuum chamber for the synchrotron light two beam operation for CHESS;
\item High resolution for low emittance optics correction and tuning;
\item Turn-by-turn readout capability for multiple bunches to support beam dynamics studies;
\item Capability for digitizing single species bunch trains with bunch spacing as small as 4~nsec and dual beam digitization for bunch trains with 14~nsec spacing.
\end{itemize}
The need for dual beam operation of the system places a unique constraint on the CESR BPM specifications.  Since the relative arrival time of the bunches from the two beams varies widely from location to location around the ring, standard RF processing techniques to optimize resolution and minimize timing sensitivity cannot be applied to the full system. As a result, the CESR design utilizes peak sampling with a high bandwidth digitizer and incorporates hardware and software design features to optimize the system timing performance.\\

\begin{table}[h]
    \centering
    \caption{\label{tab:BPMRequirements} CESR BPM Module Requirements\cite{JINST12:T09005}}
    \vspace*{1 ex}
    \begin{tabular}{ll}
        \hline\hline
        Parameter & Specification\\
        \hline
        Front End Bandwidth (for 4 ns bunch trains) & 500 MHz \\
        Absolute Position Accuracy (long term) & 100 $\mu$m \\
        Single Shot Position Resolution & 10 $\mu$m \\
        Differential Position Accuracy & 10 $\mu$m \\
        Channel-To-Channel Sampling Time Accuracy & 10 ps \\
        \hline\hline
    \end{tabular}
\end{table}

\subsection{System Implementation}
The CESR BPM system, described elsewhere\cite{JINST12:T09005}, consists of a network of local sensors and processors.  Each location has four beam buttons arranged in a mirror symmetric fashion, providing relative amplitude signals for a processing module.  All modules share a common controls database, timing and synchronization controls, and networked data storage.  This allows for accelerator-wide coordinated measurements.\\

\section{Beam Size Monitors}

Vertical and horizontal beam size measurements are made utilizing
the xBSM and vBSM, respectively.  These systems can operate in
parallel so that the combination of vertical and horizontal beam
sizes can be acquired simultaneously.  In addition the vBSM has a
$\pi$-polarization operational mode that allows it to measure
vertical beam size.

\subsection{X-ray Beam Size Monitors}

During the conversion of CESR to operate for CesrTA~research, two
xBSMs detectors were added to CHESS x-ray extractions lines with
synchrotron radiation (SR) source points located in two Hard Bend
magnets (38~m bending radius), one for positrons and the other for
electrons.  The one-dimensional monitors produce the vertical beam
profile turn-by-turn for bunches spaced as closely as 4~nsec.  The
processing software fits the vertical beam size and vertical
position turn-by-turn allowing the averaging the vertical beam size
independent of vertical dipole motion of the beam.  Beam size
measurements may be acquired on the same turns for each bunch in
trains of bunches with arbitrary spacings.  This system is described
in detail elsewhere.\cite{NIMA748:96to125, NIMA798:127to134}

\subsection{Visible Light Beam Size Monitor}
\label{ssec:cesr_conversion.beam_instr.other.vbsm}

\begin{figure}[htbp] 
   \centering
   \includegraphics[width=0.75\columnwidth]{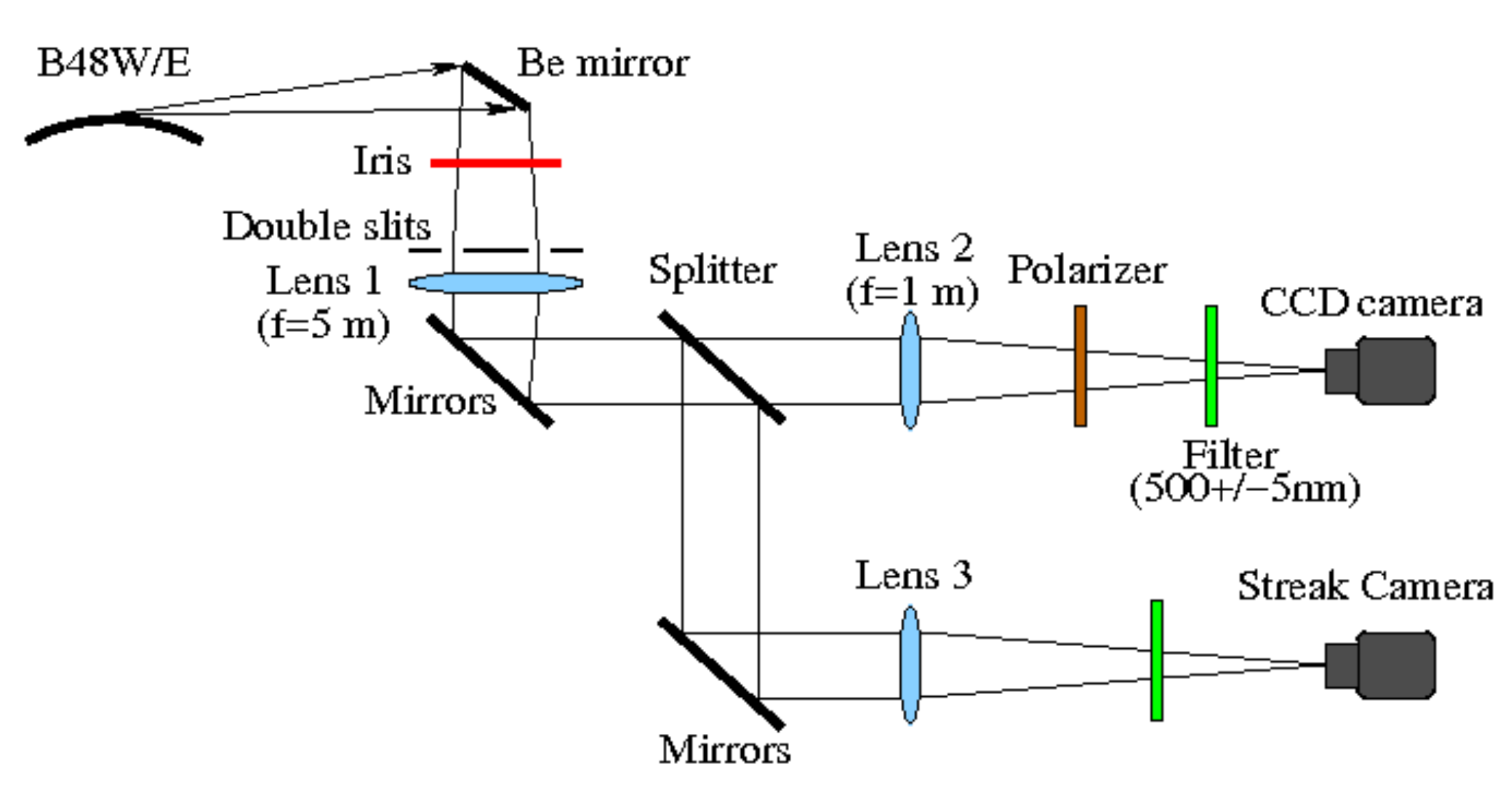}
   \caption[Overview of the {CesrTA} vBSM.]{\label{fig:L3setStreak}
   Schematic of the L3 vBSM.}
\end{figure}

\begin{figure}[htbp] 
   \centering
   \includegraphics[width=0.95\columnwidth]{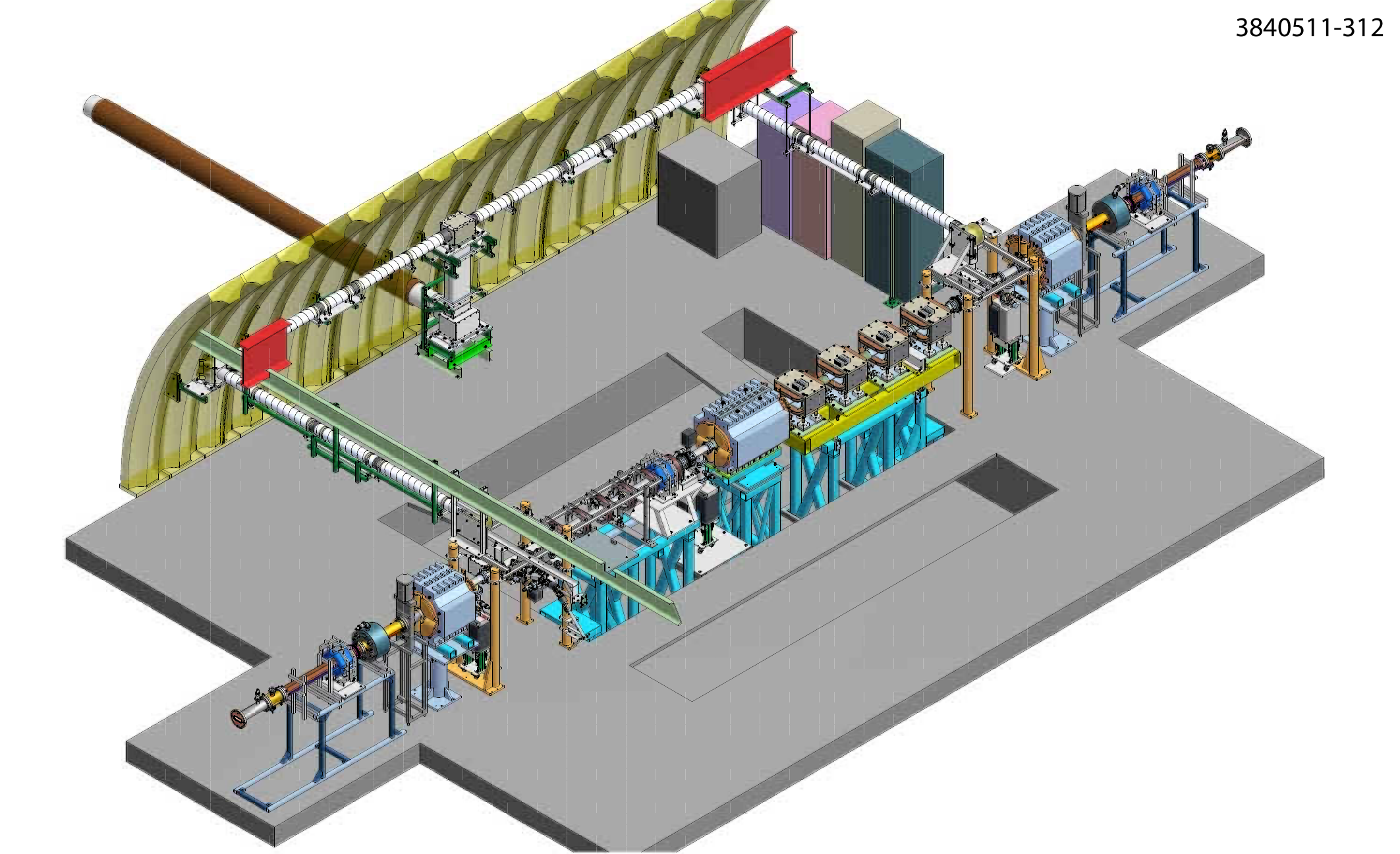}
   \caption[Final assembly drawing of the {CesrTA} beam line.]{\label{fig:L3BeamlineFinalAssembly}
   Final assembly drawing of the {CesrTA} beam line viewing L3 from above and the North and East.  The pipes coming off of the CESR beam line to the south, transport the light from the vBSM's mounted just to the North (center of straight section) of the Q48W and Q48E quadrupoles (the outermost large blue magnets.) }
\end{figure}

The two visible light beam size monitor ports were installed in the
L3 straight section of the storage ring.  An overview of the vBSM
system is shown in Figure~\ref{fig:L3setStreak}. The vBSM ports are
placed symmetrically at east and west ends of the straight section
in order to image visible synchrotron light from the electron and
positron beams, respectively. The visible SR from bending magnets
(140~m bending radius) is reflected by a Beryllium mirror located
inside the vacuum chamber, which directs the light out of the
chamber through a vacuum window and into an optics box. On exiting
the vacuum window, the SR photons pass through an iris with an
adjustable aperture into the optics box. The optics box contains
several sets of double slits with different slit spacings and
orientations for interferometric measurement of vertical or
horizontal beam size. The slits are followed by a focusing lens. The
light is then reflected by multiple mirrors, overhead, across the
tunnel and eventually through the wall from the accelerator tunnel
to an optical table in an experimental hall.  At this point the
light has traveled 27~m from the source in the bend magnet. The path
of the light, from vacuum window through tunnel wall, is indicated
in Figure~\ref{fig:L3BeamlineFinalAssembly}. On the optical table in
the experimental hall, the SR light passes through a second lens, a
polarizer, a bandpass filter and is finally incident on a CCD
camera.  A beam splitter can be placed in the path to direct a
fraction of the light into a streak camera in order to measure the
bunch length. The vBSM system is explained in detail in
\cite{NIMA703:80to90}.

\section{Tune and Motion Detection}

The variation of the tunes of individual bunches within trains of
bunches or of a witness bunch, following a train of bunches, carries
information about the electron cloud density.  Different methods
have been employed to measure the  tunes of the bunches during the
beam dynamics studies.

\subsection{Tune and Motion Detection For Multiple Bunches By Turn-by-Turn Trajectory Measurements}

A simple method for determining the tunes for each bunch in a train
of bunches is to use a subset of the complete number of CBPM modules
to measure the beam position turn-by-turn for each bunch. The data
is read out from the CBPM modules and written into a raw data file.
Each BPM's position data is then analyzed offline by performing a
Fourier transform, which yields the spectral lines of the beam's
transverse motion.  This method is most often used in conjunction
with a kicker that deflects all of the bunches within the train on a
single pass (see Section 5.2).


\subsection{Tune and Motion Detection By Single Bunch Spectrum Measurements}

\begin{figure}[htbp] 
   \centering
   \includegraphics[width=0.6\columnwidth]{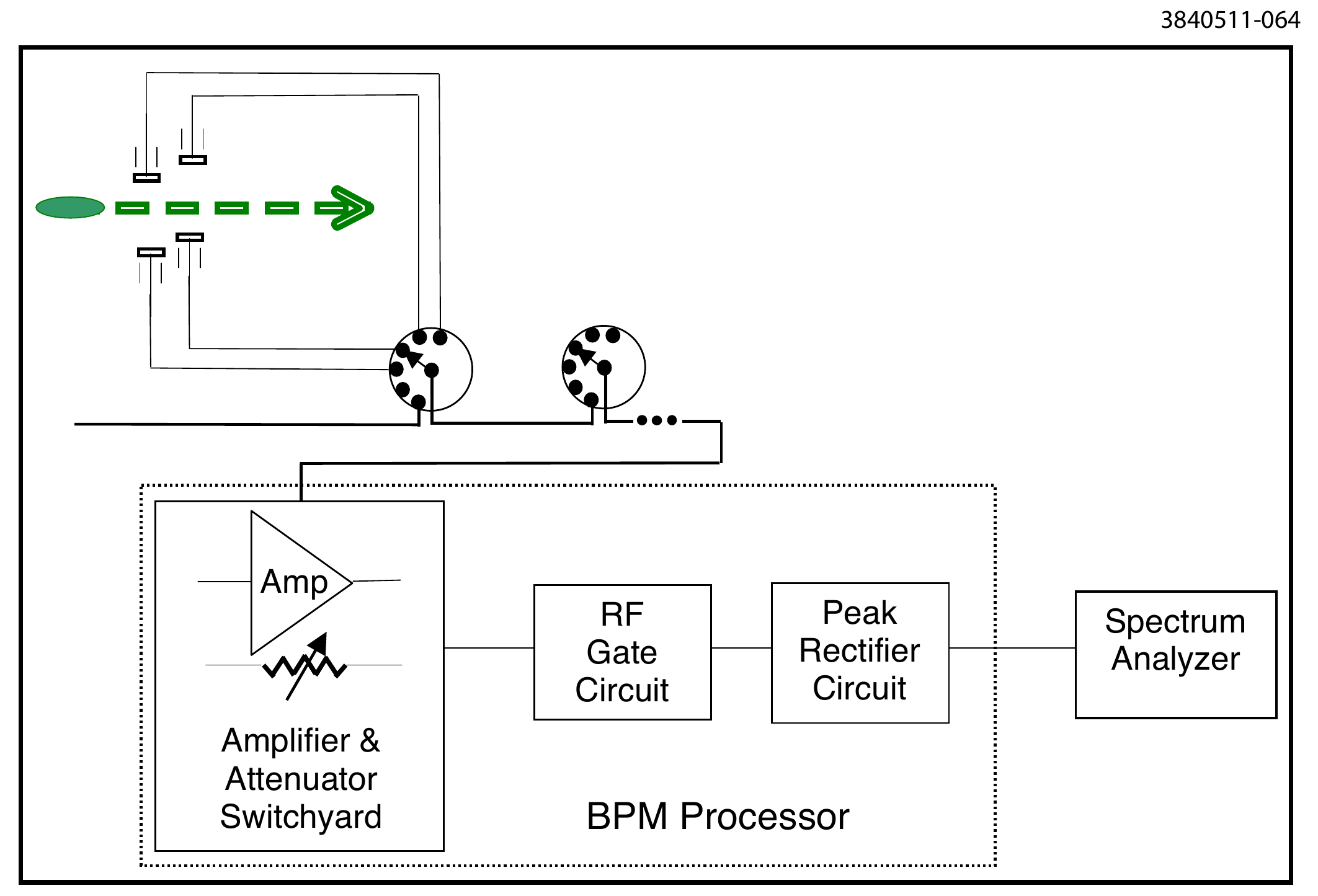}
   \caption[Relay BPM tune receiver]{\label{fig:RelayBpmTuneReceiver}
   Block diagram of a betatron tune receiver using the relay BPM system. }
\end{figure}

A second method for detecting the tunes of a single bunch within a
train of bunches is shown in the block diagram in
Figure~\ref{fig:RelayBpmTuneReceiver}.  This detection method makes
use of one of a few BPM detectors, which are still connected to
CESR's original relay-based BPM system processors.  The signal from
one BPM button is routed via coaxial relays to one of the analog
processors, where fixed gain amplifiers and/or attenuators may be
inserted in the signal path to maintain the peak signal level within
a factor of five over a wide range of currents.  After the gain
adjustment the signal passes along to an RF gating circuit, which is
triggered by signals from CESR's fast timing system.  This allows
the gating of the signal from a single bunch, sending it to a peak
rectifier circuit (with approximately a 700~MHz bandwidth) and then
routing its video output to a spectrum analyzer in the Control Room.
This hardware has a much better sensitivity than the CBPM; it is the
only position monitoring system at CESR capable of observing
head-tail motion of bunches.

\begin{figure}[htbp] 
   \centering
   \includegraphics[width=0.6\columnwidth]{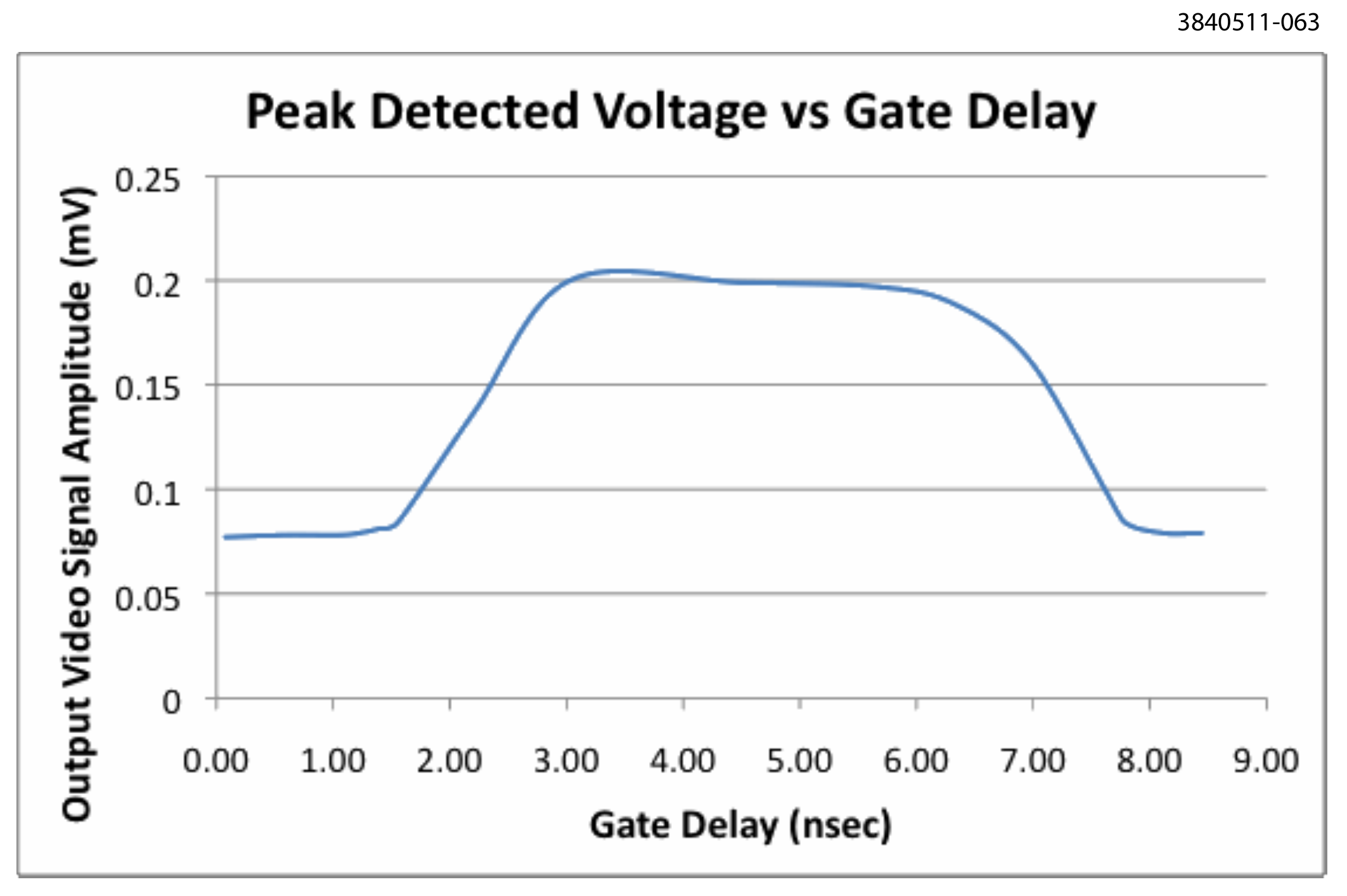}
   \caption[Relay BPM tune receiver timing aperture - Raw signal level]{\label{fig:RelayBpmTimingAperture-RawSignal}
   Relay BPM processor's gate timing aperture as measured with the video signal in the Control Room having a 77~mV~DC offset due to the peak rectifier circuit. }
\end{figure}

\begin{figure}[htbp] 
   \centering
   \includegraphics[width=0.6\columnwidth]{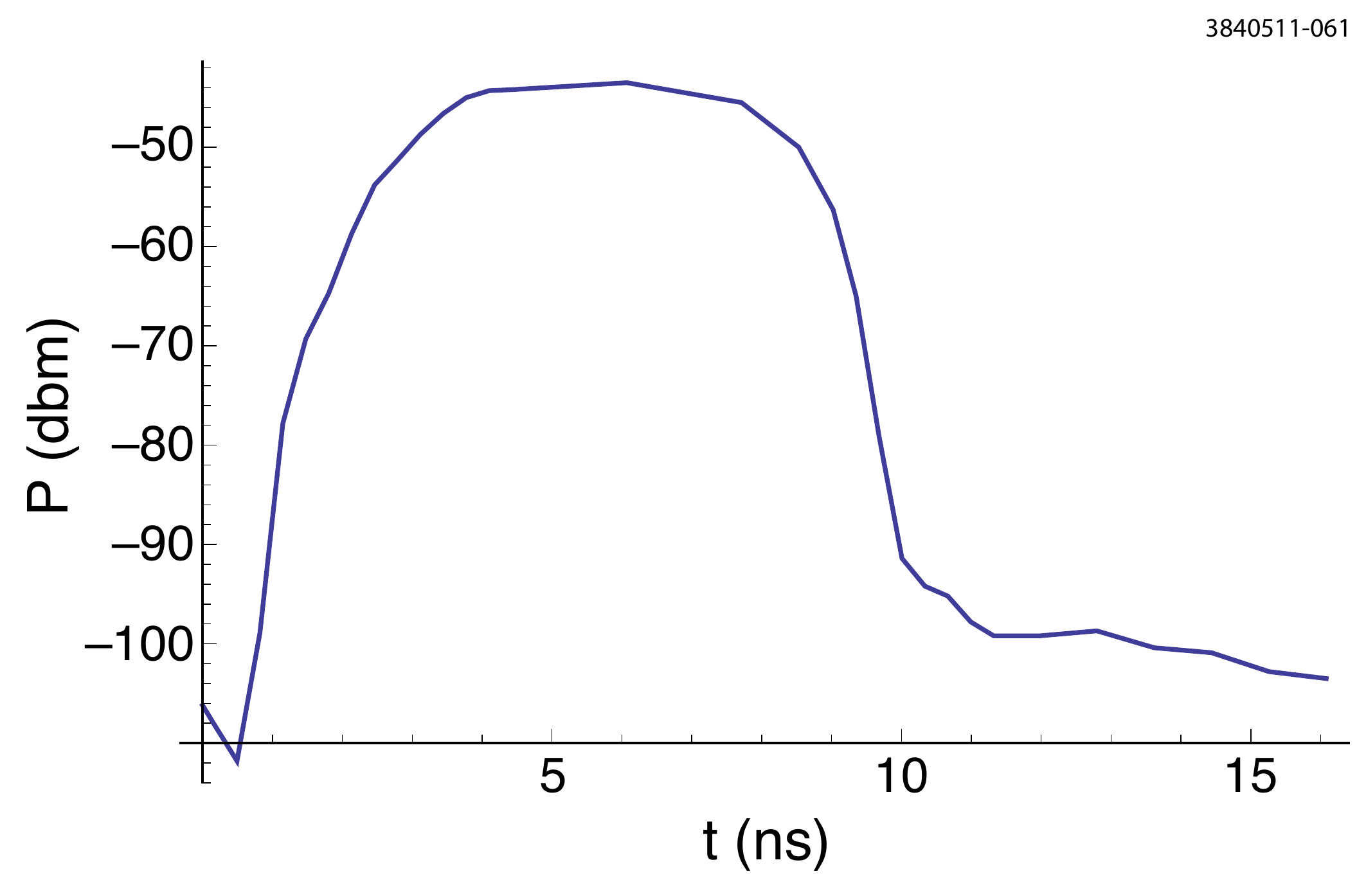}
   \caption[Relay BPM tune receiver timing aperture - Vertical betatron signal level]{\label{fig:RelayBpmTimingAperture-BetatronTuneSignal}
   Relay BPM processor's gate timing aperture as measured by driving a single bunch vertically and measuring its response vs. gate delay. }
\end{figure}

The timing aperture for the gating circuit was measured by sweeping
the gate delay for the signal coming from a single bunch to observe
the signal amplitude vs. gate delay and the results are displayed in
Figure~\ref{fig:RelayBpmTimingAperture-RawSignal}.  A second method
for observing the signal crosstalk between bunches is seen in
Figure~\ref{fig:RelayBpmTimingAperture-BetatronTuneSignal}.  This
plot is obtained by shaking the beam vertically and observing the
spectrum analyzer's signal amplitude as a function of gate delay.
This second observation gives the base timing aperture as 7.5~nsec
wide, giving more than 20~dB isolation of the signal crosstalk from
adjacent 4~nsec-spaced bunches and a signal isolation of greater
than 50~dB for 14~nsec-spaced bunches.

\subsection{Single Bunch Tune Measurements To Establish the Operating Point in the Tune Plane}

\begin{figure}[htbp] 
   \centering
   \includegraphics[width=0.4\columnwidth]{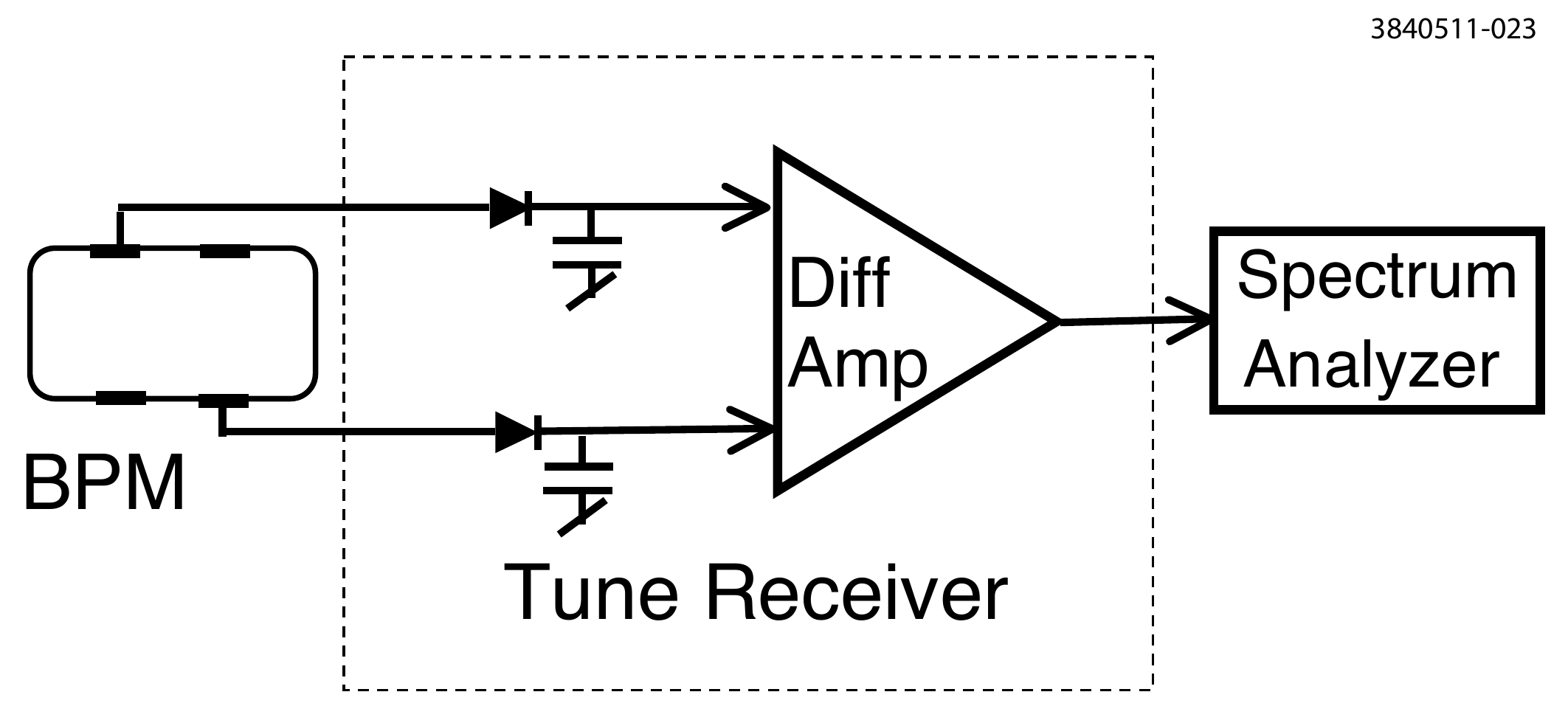}
   \caption[Narrow bandwidth BPM tune receiver]{\label{fig:DedicatedBpmTuneReceiver}
   Simple block diagram of a narrow-band tune receiver, shown in the mode where it can detect both the horizontal and vertical betatron tunes. }
\end{figure}

The initial setup of the storage ring operating point in the tune
plane at the beginning of a set of dynamics measurements requires
the determination of the betatron and synchrotron tunes, which are
routinely performed by measurements with a single stored bunch. The
betatron tune instrumentation configuration, which is capable of
detecting the beam's tune in both planes, is shown in the block
diagram in Figure~\ref{fig:DedicatedBpmTuneReceiver}.  In this mode
the single bunch is excited with the relatively narrow bandwidth
shaker magnets (see Section 5.3) and detected with a swept spectrum
analyzer.  For synchrotron tune measurements the RF accelerator
system's phase command is modulated and the energy oscillation is
detected using the orbit's dispersion in one BPM detector.

\section{Beam Excitation}

To measure the tune spectra of bunches it is necessary to observe
them undergoing coherent motion.  In some cases their
self-excitation is sufficient for a good tune measurement, but in
other cases the beam must be driven with some type of dipole kicker
to cause the beam to undergo centroid motion.  There are three types
of dipole kickers used in CESR.

\subsection{Pinger Magnets}

\begin{figure}[htbp] 
   \centering
   \includegraphics[width=0.6\columnwidth]{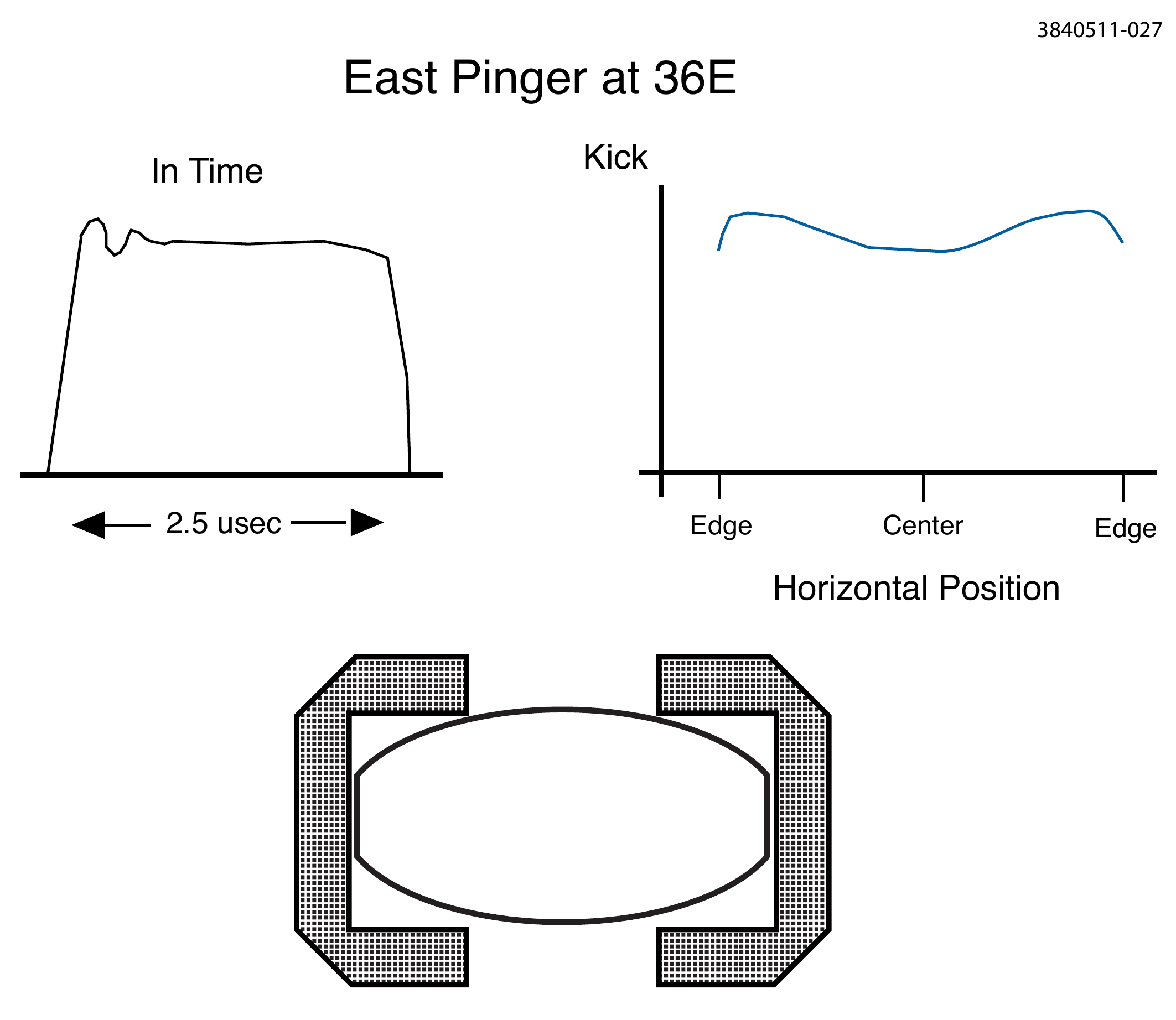}
   \caption[Pinger beam deflection magnet]{\label{fig:EastPinger}
   Horizontal pinger. This is a pulsed ferrite magnet surrounding a metallized coated ceramic vacuum chamber, which provides deflection to the beam with a single turn's duration.   The upper two plots display qualitatively the deflection to the beam as a function of time (on the left) and spatially across the vacuum chamber's aperture (on the right.)}
\end{figure}

A pinger magnet is used to create a single impulse deflection for
the beam. There are three pingers installed in CESR: two are
horizontal and one is vertical.  A horizontal pinger is shown in
Figure~\ref{fig:EastPinger} and is a pair of single-turn
ferrite-core magnets, which surround a Kovar-coated ceramic vacuum
chamber.  The horizontal pingers are excited using a thyratron with
an approximately square pulse, having a flattop region about
2~$\mu$sec long, where the circulation time for CESR is
2.56~$\mu$sec.  This is more than long enough to excite all bunches
in one train with the same deflection angle.   The pulse shape for
the vertical pinger has a different waveform, since the magnet is
constructed in a ferrite box structure with two 2-turn windings; the
magnet is driven with a half sine-wave pulse of approximately
2.5~$\mu$sec duration.  The pingers can be triggered via CESR's Fast
Timing System at repetition rates up to 60~Hz and the triggers can
be synchronized with the CBPM turn-by-turn and bunch-by-bunch data
acquisition.  Because of its half sine-wave shape, for the
excitation of a train of bunches the vertical pinger is timed to
have the bunches in a train arrive at the time for them to straddle
the peak of the pinger's deflection.

\subsection{Stripline Kickers}

\begin{figure}[htbp] 
   \centering
   \includegraphics[width=0.6\columnwidth]{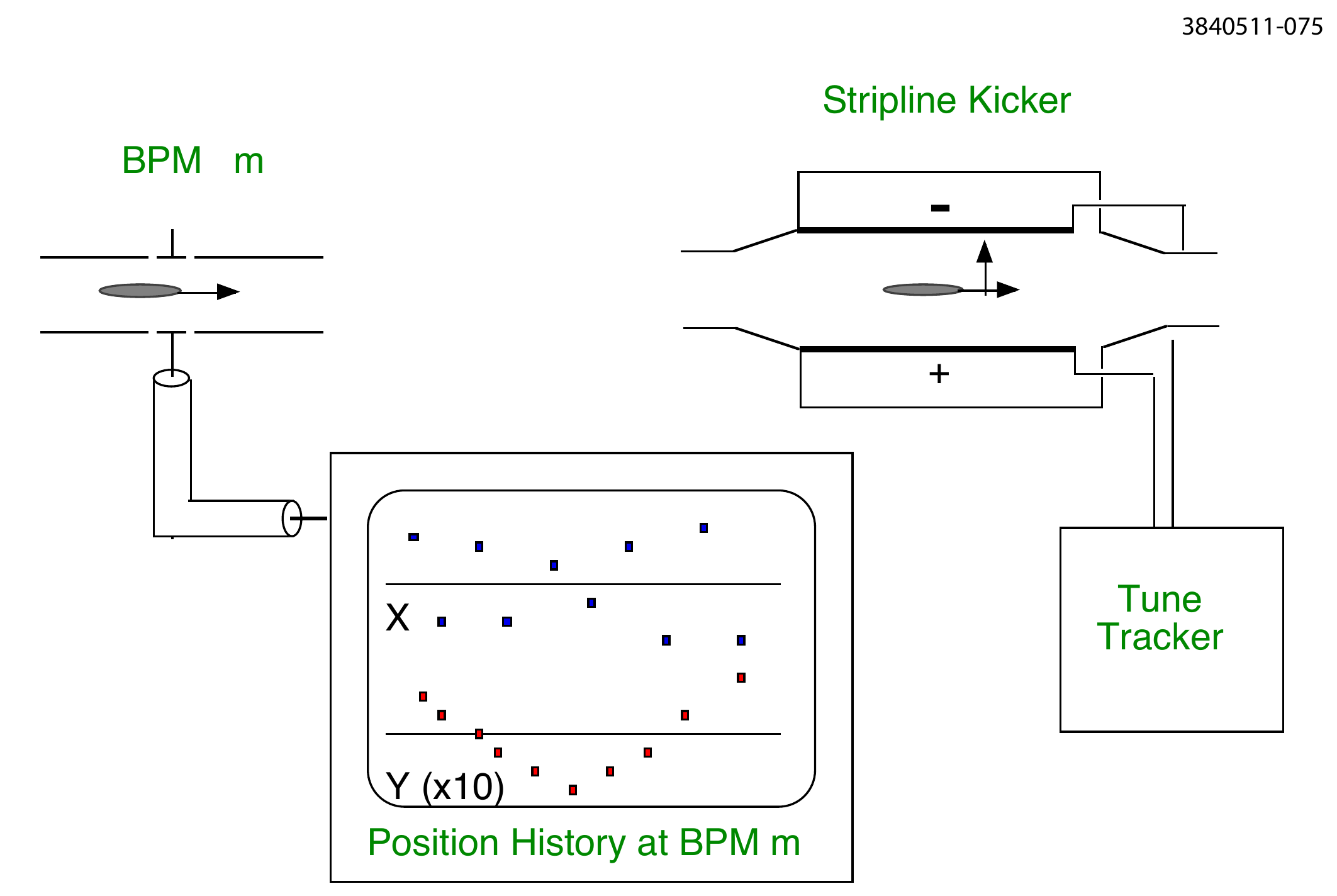}
   \caption[Stripline kicker]{\label{fig:StriplineKicker}
   Dipole excitation of a bunch utilizing a stripline kicker, which has two plates that are driven differentially to deflect the bunch. }
\end{figure}

The second type of deflection element is a stripline kicker, which
is utilized for beam dynamics measurements in the configuration
shown in Figure~\ref{fig:StriplineKicker}.  There are two stripline
kickers installed in CESR, one horizontal and one vertical, and
serve as are the deflectors for the transverse dipole bunch-by-bunch
beam stabilizing feedback systems for the ring.  The kickers are
shorted at one end, with a transit time of 3.5~nsec, and are excited
with 200~W-150~MHz bandwidth RF amplifiers. The kickers have an
impedance of 50~$\Omega$ and are constructed from OFHC (oxygen free,
high thermal conductivity) copper sheet.  As a part of the
transverse feedback system for 14~nsec-spaced bunches, the
amplifiers are modulated with 14~nsec single period sine-wave,
producing an approximately constant ($\pm$5~$\%$) deflection to the
beam for about 3.5~nsec.  The crosstalk of this stripline kicker to
adjacent bunches is less than -40~dB for 14~nsec-spaced bunches.
Each feedback modulator has an external modulation input and when it
is enabled, the input will allow the deflection of any combination
of 14~nsec-spaced bunches.  For beam dynamics measurements, the
stripline kickers are most often used to deflect individual bunches
within the train.  For detailed tune shift measurements, another
method excites the first bunch at the same time one of the later
bunches is being excited; this allows accurate accounting for the
difference in tunes between the lead bunch and the measured bunch
when drifts in the accelerator tunes are present.

\subsection{Shaker Magnets}

For completeness there is a third type of deflection component in
the storage ring.  This is low-frequency shaker magnet, a multi-turn
coil wound around a H-frame ferrite core surrounding a metalized
coated ceramic vacuum chamber.  Although this shaker magnet does not
have sufficient bandwidth to excite motion of individual bunches and
is therefore not in use during beam dynamics measurements of trains
of bunches, it is still important for the detection of the tunes,
when CESR conditions are re-established at the beginning of each
extended measurement period.

\section{Digital Tune Tracker}

\begin{figure}[htbp] 
   \centering
   \includegraphics[width=0.75\columnwidth]{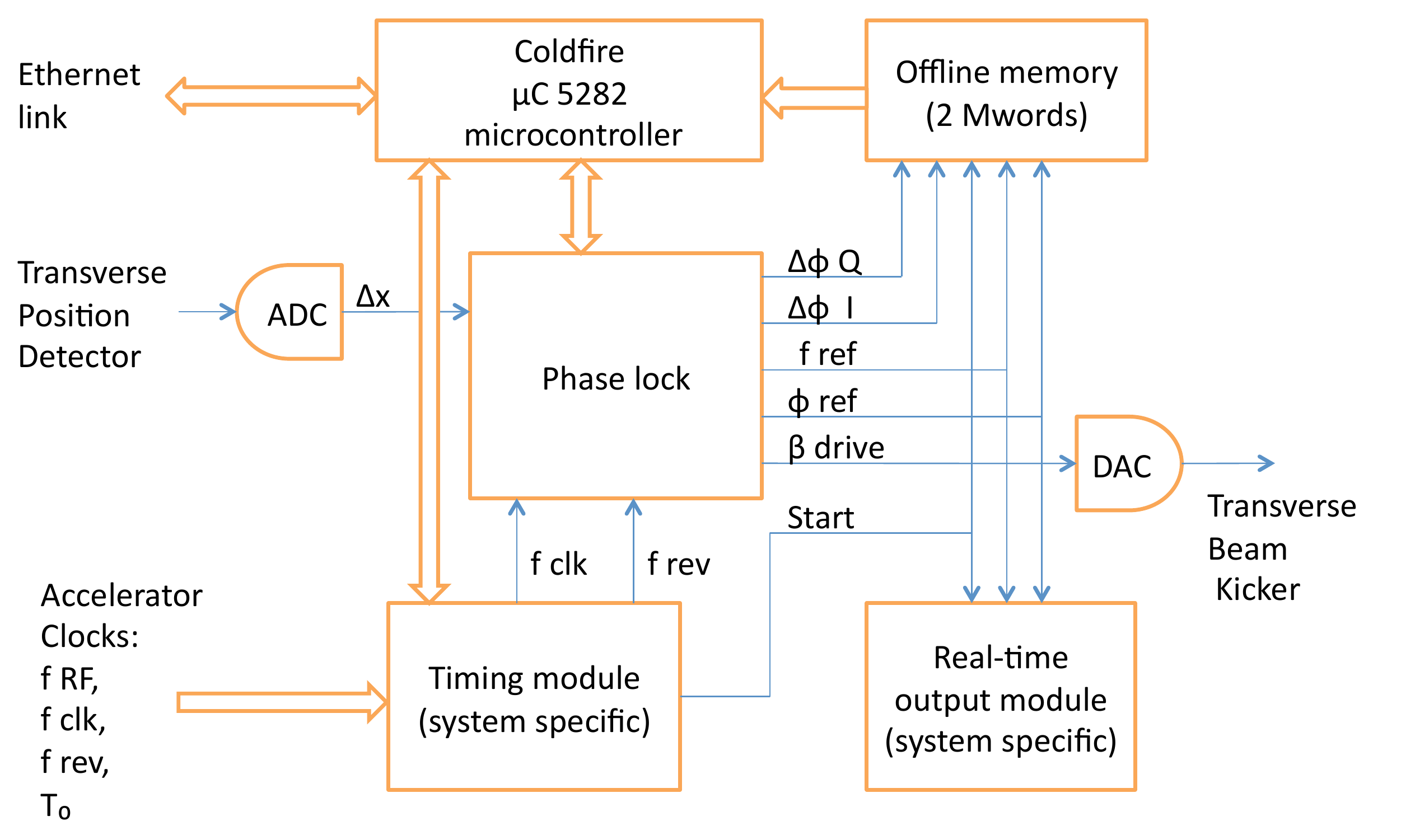}
   \caption[Tune Tracker]{\label{fig:tune_tracker_block}
   Overall block diagram of the tune tracker.  The phase-lock block diagram is detailed in Figure~\ref{fig:tune_tracker_phase}}
\end{figure}

Numerous storage ring diagnostic procedures require synchronous
excitation of beam motion. Some examples are the lattice betatron
phase advance measurement\cite{PRSTAB3:092801} and BPM
button-by-button gain calibrations\cite{PRSTAB13:092802}, which
involve synchronous detection of the driven betatron motion. In CESR
the transverse tunes can vary continuously by several times their
natural width. Hence, synchronous beam excitation is impossible
without active feedback control. The digital tune
tracker\cite{PAC11:MOP215} consists of a direct digital frequency
synthesizer, which drives the beam through a transverse kicker, and
is phase locked to the detected betatron signal from a BPM. This
ensures synchronous excitation and, by setting the correct locking
phase, the excitation can be adjusted to be at the betatron tune's
resonant peak. The fully digital signal detection allows selecting a
single bunch within a long train to be synchronously driven, which
permits lattice diagnostics to be performed including collective
effects.  An overall block diagram of the tune tracker is shown in
Figure ~\ref{fig:tune_tracker_block}.

The tune tracker operates at a clock frequency of 71.4 MHz,
corresponding to 14~nsec bunch spacing.  This clock is used for beam
sampling, filtering, and synthesis of the betatron drive signal.
Since only a single bunch is to be used for phase locking, the
instrument can select any bunch from within any pattern of stored
bunches.

\begin{figure}[htbp] 
   \centering
   \includegraphics[width=0.75\columnwidth]{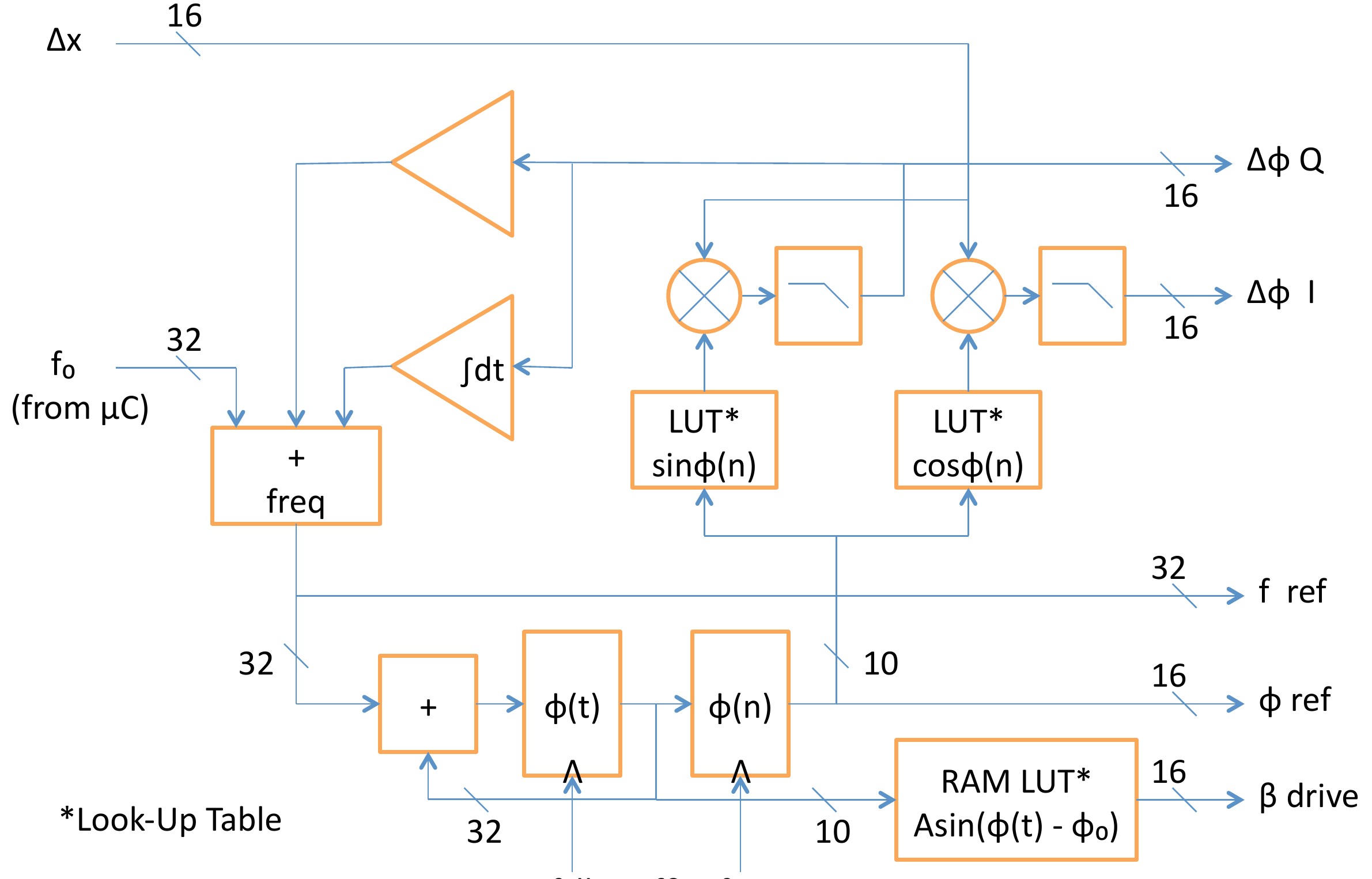}
   \caption[Tune Tracker]{\label{fig:tune_tracker_phase}
   Diagram of the phase-lock block, which was shown as part of Figure~\ref{fig:tune_tracker_block}.}
\end{figure}

The position signal is taken from a set of microstripline
electrodes, which are separated into amplitude and displacement
signals with a network of sum and difference combiners. The
difference signal is digitized with a 10 bit ADC, which is timed to
peak signal amplitude, and the signal from the selected bunch is
latched for one turn. As shown in
Figure~\ref{fig:tune_tracker_phase} the latched amplitude signal is
digitally mixed at 71.4~MHz with two square wave representations of
the betatron drive signal at quadrature phases. This produces a
vector representation of the phase difference between the
synthesized betatron drive and the actual betatron motion of the
beam. The betatron clock is represented as square waves to eliminate
the need for real-time multiplication. The demodulated position
signals are filtered in a pair of single pole infinite impulse
response filters. One of the filtered signals is used to represent
betatron phase error and the other is only used to reconstruct
signal amplitude.

The direct digital synthesizer (DDS) consists of a phase register,
which is incremented by the frequency command at the 71.4~MHz clock
rate, a sinusoidal lookup table implemented in a high-speed cache
RAM and a 14~bit DAC. Adjustments of the drive phase and amplitude
are effected by changing the contents of the RAM with the 14~bit
output resolution giving sufficient dynamic range for all
applications without the need for analog attenuation. The betatron
drive signal is coupled to the beam via the feedback kicker, which
allows the isolated drive of a single bunch in the 14~nsec spacing
configuration. For 4~nec-bunch spacings, there is -20~dB crosstalk
of the drive signal to bunches adjacent to the one selected for
phase locking, due to the modulator waveform shape and the 3.5~nsec
stripline kicker length.

The phase-locked loop requires a proportional channel and an
integrating channel. The proportional channel shifts the betatron
frequency command by an amount proportional to the phase error. This
is necessary to maintain loop stability and to give the loop
sufficient agility to track the tune fluctuations of the storage
ring in real time. The integrating channel increments the frequency
command on every revolution by an amount proportional to phase
error. This is necessary to bring the phase error to zero and thus
provide a stable phase reference for lattice measurements.

The DDS phase register value is latched once per accelerator
revolution and the phase is sent by a parallel digital link to the
clock modulator for the CBPM system, \cite{JINST12:T09005,
IPAC10:MOPE089}. The BPM clock modulator imposes both the vertical
and horizontal phase values from the two tune trackers on the BPM
clock using a pulse width modulation system. The individual BPM
modules then extract the phase values and use them to reconstruct
the drive signals (typically at one of the three normal mode dipole
frequencies: horizontal and vertical betatron modes or the
synchrotron oscillation mode), which is used to synchronously detect
corresponding phase at each BPM station. The synchronous phase
measurement is used to determine the phase advance between BPM
stations, and hence the phase function of the entire lattice, while
the relative phase of horizontal and vertical motion at each BPM is
used to extract coupling information\cite{PRSTAB3:092801}.

The operating configurations of the tune trackers, including center
frequencies, gains, and filter settings, are saved and restored
along with the storage ring configuration. This gives a high
probability of a successful phase lock with minimal adjustment. The
signal acquisition and locking functions can be operated through a
graphic user interface and the same functions can be executed
automatically by other system processes using control system
subroutines.

The tune tracker can initially acquire a betatron signal by sweeping
the drive frequency through a band, typically 20~kHz wide, and
recording betatron amplitude and phase error relative to the DDS. A
fit of center frequency, center phase, peak width, and peak
amplitude is then automatically computed to match a Lorentzian
resonance model, where the phase shifts by~$\pi$~as the excitation
frequency sweeps across resonance.

\begin{figure}[htbp] 
   \centering
   \includegraphics[width=0.75\columnwidth]{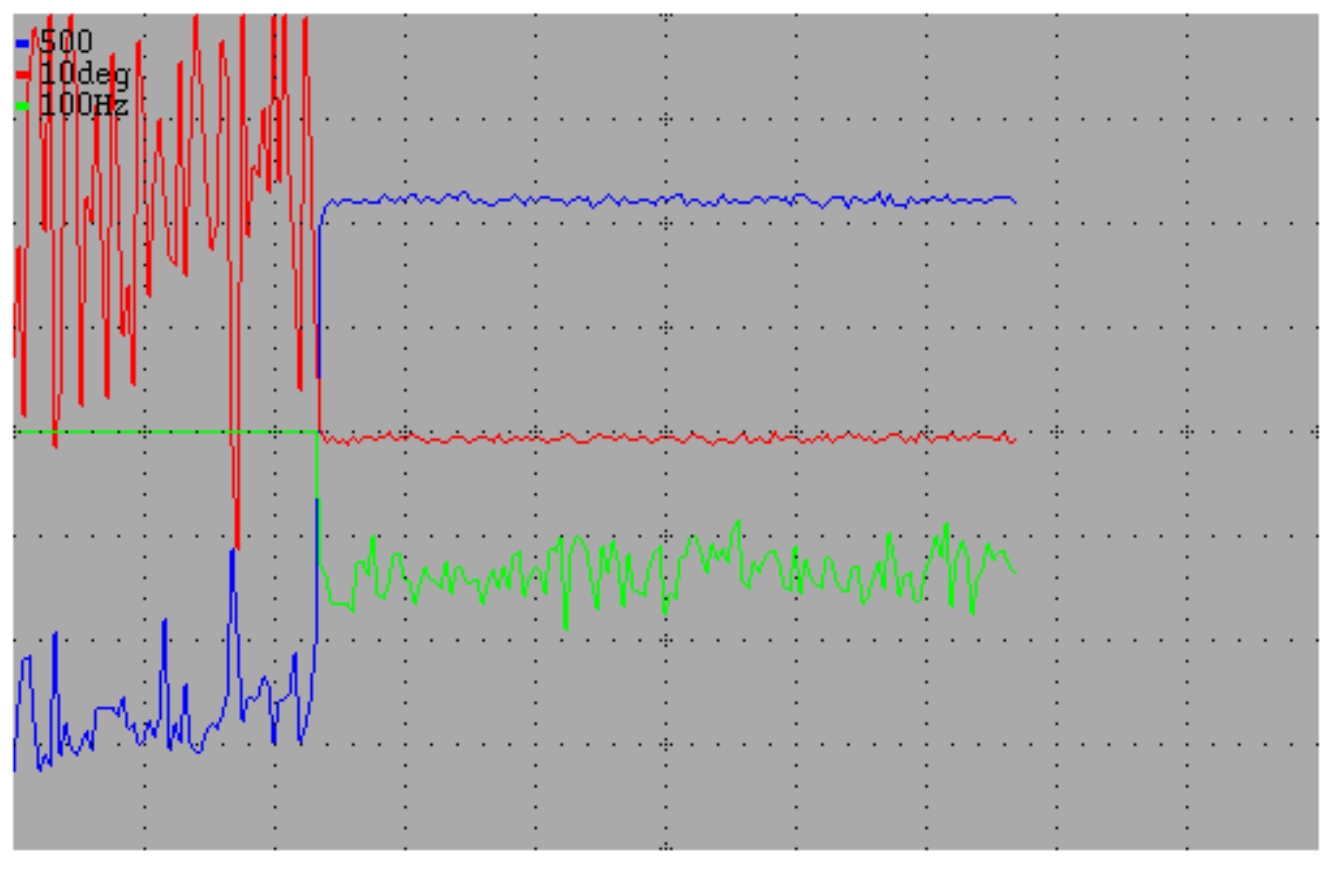}
   \caption[Tune Tracker]{\label{fig:time_sweep}
   Time sweep showing the acquisition of the lock condition for the tune tracker.  The blue trace is the betatron amplitude (arbitrary units). The red trace is the betatron phase error (10 degrees per division.)  The green trace is the direct digital synthesizer frequency offset (100~Hz per division.)  The time scale is approximately 0.4~sec per division.  Note that the tune tracker is following the frequency shifts of the betatron tune after the lock condition is established.}
\end{figure}

Only a rough fit is possible to the tune of an unlocked beam because of the tune noise of the storage ring, which is approximately a few hundred Hertz. Acquisition of phase lock, seen in Figure~\ref{fig:time_sweep}, is accompanied by a large increase in betatron amplitude. Once phase lock is established, a fine fit to the resonance can be done by sweeping the betatron drive phase, and recording betatron amplitude and drive frequency. A fit of the same four parameters is automatically performed using the Lorentzian model by the same method described above. This consists mainly of fitting the amplitude to a cosine function, and gives the closest possible approach to the resonance peak. 

\section{Beam Stabilizing Feedback Systems}

Three Dimtel iGp-1281F signal processor systems have been added to
the CESR ring to supplement existing feedback systems. Prior to this
CESR operated exclusively with 14~nsec, multiple bunch, turn-by-turn
feedback\cite{PAC01:TPAH006}. The 14~nsec system was designed for
and constrained by having electrons and positrons in the ring
simultaneously. The Dimtel systems add faster processing capability
to the same detection and kicker hardware to provide independent
feedback for all bunches in a single electron or positron beam with
spacings down to 4~nsec. This gives CESR the flexibility to
transversely and longitudinally stabilize bunch trains with a bunch
spacing of any integer multiple of 2~nsec greater than 4~nsec.

\subsection{14~nsec Feedback System}

\begin{figure}[htbp] 
   \centering
   \includegraphics[width=0.95\columnwidth]{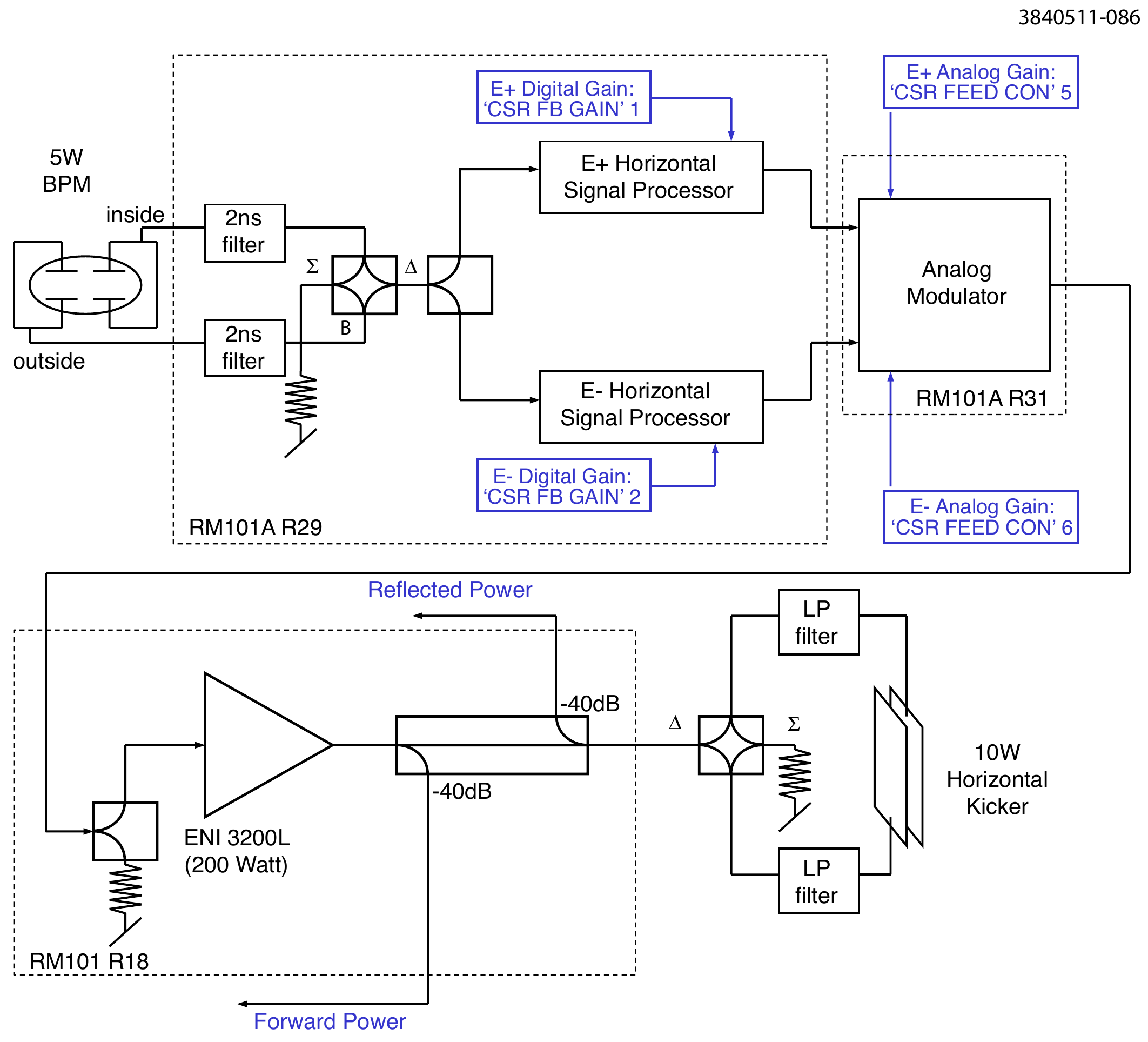}
   \caption[14~nsec horizontal feedback system]{\label{fig:14nsFbHorzSys}
   System overview of the horizontal 14~nsec feedback.}
\end{figure}

The 14~nsec feedback system\cite{PAC01:TPAH006} uses a standard set
of four CESR beam position buttons as the input.  Pairs of button
signals are connected to hybrid combiners which output both the sum
and difference of the two signals. The vertical feedback system uses
the signals from combining the top and bottom button signals
together, thus yielding a difference signal which is sensitive to
vertical position. Likewise, the horizontal system uses the
difference of the combined inner and outer button signals. The sum
of all four button signals is used as the longitudinal system input.
The horizontal system block diagram is nearly identical to the
vertical system block diagram and can be seen in
Figure~\ref{fig:14nsFbHorzSys}.

Signals are acquired at 71.4~MHz by a direct sampling 10~bit analog
to digital converter (ADC). This rate allows sampling at the
traditional CESR bunch spacing of 14~nsec. Longitudinal signals are
processed with displacement or phase detection, sample and hold,
displacement or phase offset correction, and filtering circuits.
Transverse signals are filtered then sampled in a digital signal
processor (DSP.) The output is the beam error signal represented by
a $\pm$500~mV analog signal that is updated every
14~nsec\cite{PAC95:RAE14}.

For the transverse systems, the error signal is modulated into a
bipolar pulse. The pulse is amplified by an ENI3200L 200~watt power
amplifier with bandwidth from 250~kHz to 150~MHz. The amplified
pulse is sent to a 1.16~m stripline kicker.

\begin{figure}[htbp] 
   \centering
   \includegraphics[width=0.95\columnwidth]{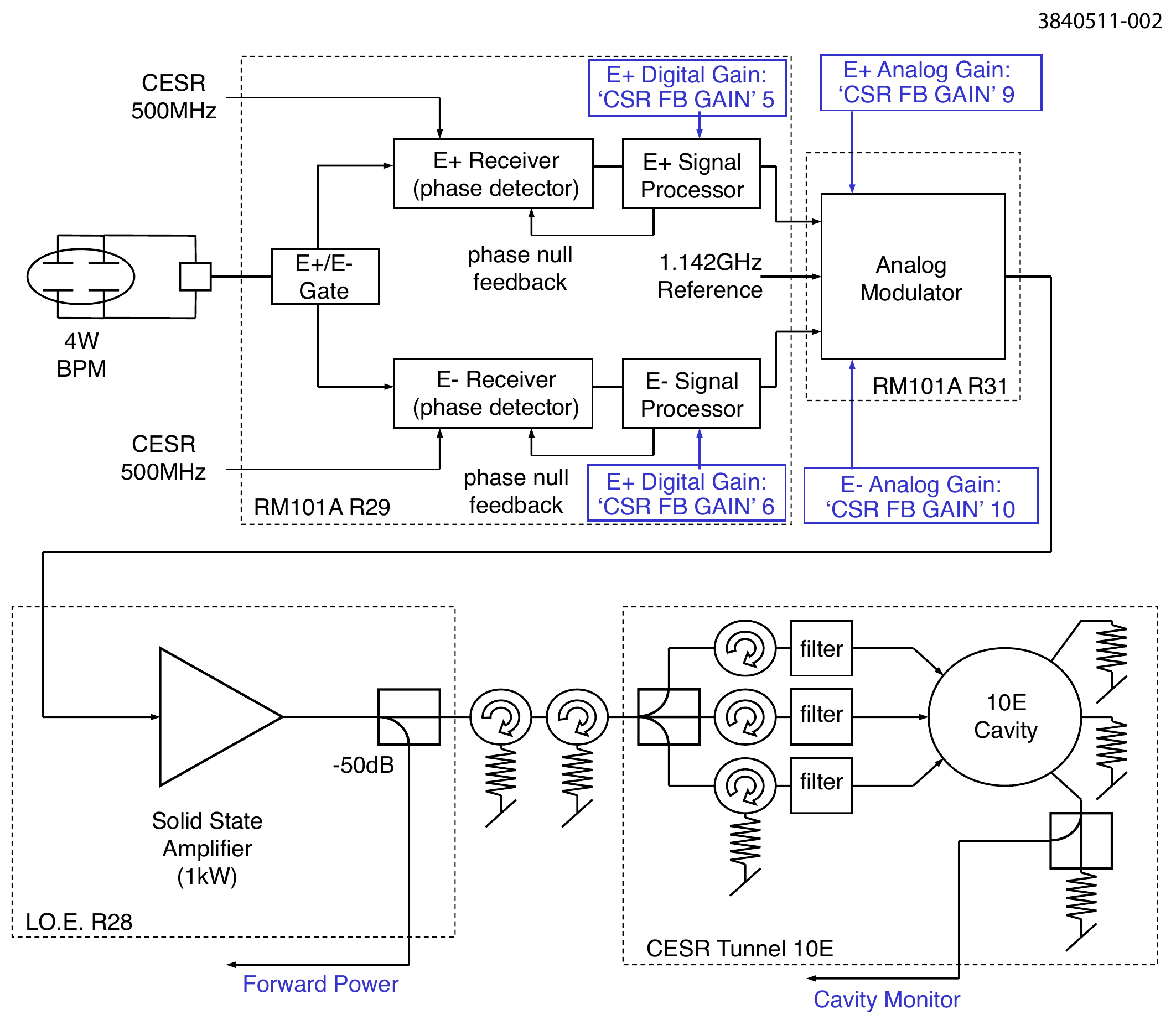}
   \caption[14~nsec longitudinal feedback system]{\label{fig:14nsFbLongSys}
   System overview of the longitudinal 14~nsec feedback.}
\end{figure}

The longitudinal error signal is amplified by a 1~kW solid state
amplifier. The signal is transmitted to the beam through an 1142~MHz
resonant frequency cavity kicker. The cavity kicker has three
coupling ports for drive and three coupling ports for load, for a
total of six ports.  The cavity has a loaded $Q$ of about 14 and a
field decay time of 3.9~nsec\cite{PAC01:TPAH006}. A block diagram
overview of the longitudinal system is in
Figure~\ref{fig:14nsFbLongSys}.

The 14~nsec feedback system is controlled through the CESR control
system multi-port memory (MPM). The MPM allows many devices to share
a memory space.  The DSPs access the shared memory for information
on the gains and delays necessary for different CESR ring
configurations.

\subsection{4 ns Feedback System}

\begin{figure}[htbp] 
   \centering
   \includegraphics[width=0.7\columnwidth]{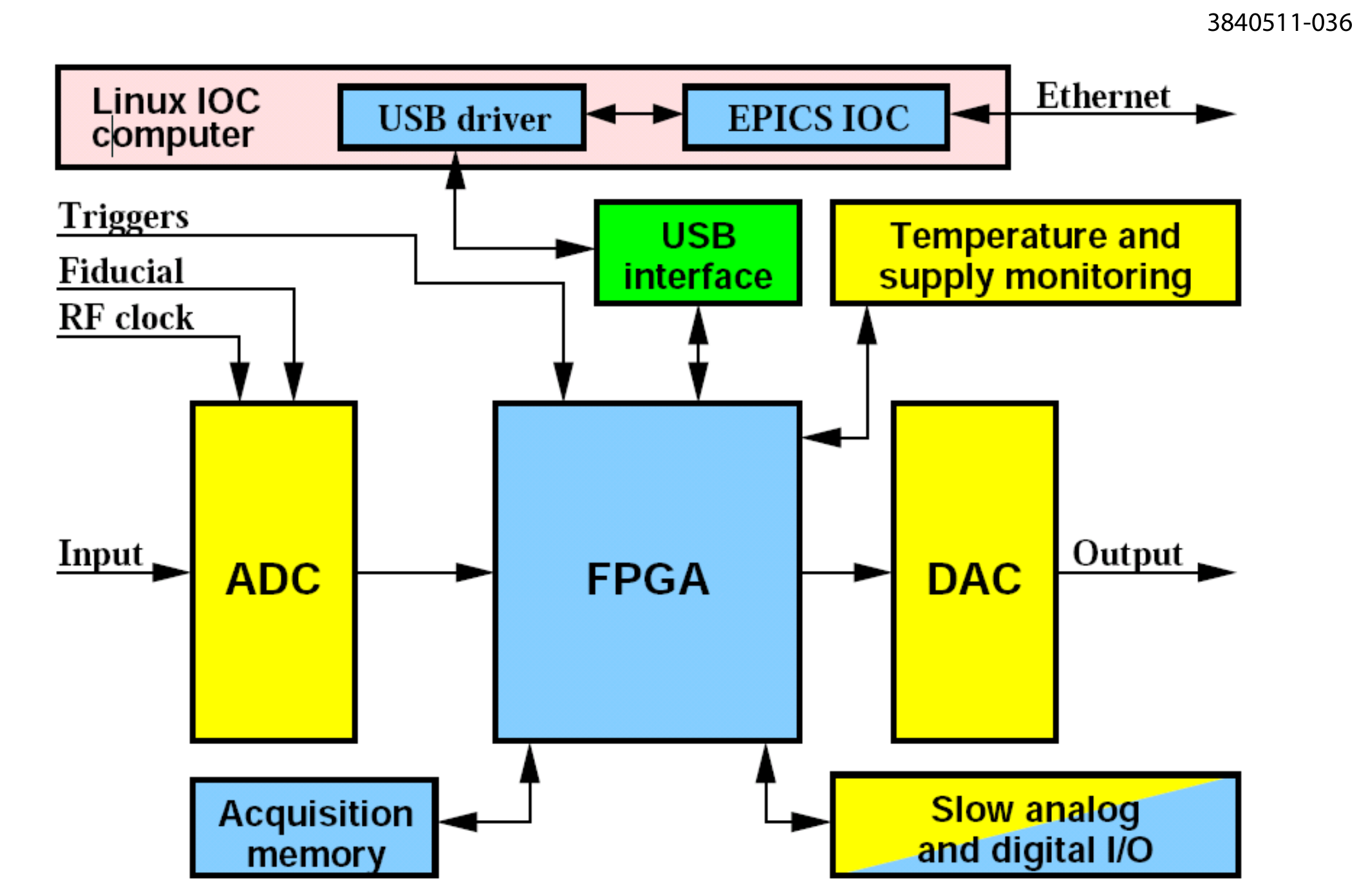}
   \caption[Dimtel iGp system overview]{\label{fig:4nsFbSys}
   System overview of the Dimtel iGp 4~nsec signal processor.}
\end{figure}

The pre-existing feedback system described above was augmented to
provide stability for bunches spaced down to 4~nsec by adding three
Dimtel iGp-1281F signal processors\cite{Dimtel:REF}, one each for
horizontal, vertical and longitudinal dipole modes of oscillation.
The processors use the same beam input signals as the 14~nsec
feedback system as well as the same stripline and cavity kickers.  A
block diagram of the iGp is in Figure~\ref{fig:4nsFbSys}.

The new signal processors operate at 500~MHz and acquire the input
signals with 1.26~GHz bandwidth with high speed 8~bit ADCs.  The
signals are processed by a dual-port memory field programmable gate
array (FPGA) for control computations including applying a finite
impulse response (FIR) filter with up to 16~taps.  A high speed DAC
drives the output signal with 12~bit resolution, a rise time under
250~psec and a fall time under 350~ps.

The FPGA is controlled by an embedded EPICS input-output controller
(IOC).  The IOC is connected to the FPGA by universal serial bus
(USB) and to the CESR control system via
Ethernet\cite{DIMTEL2009:IGP1281F:Man17}.  Computing scripts and
display screens on the CESR control system are used to control the
parameters of the signal processor and to acquire and analyze data.
An additional software EPICS IOC running on the CESR control system
and a service program provide an interface from the CESR MPM to the
iGp devices.

Completely configurable bunch control allows the setting of any
combination of the 1281 available 2~nsec RF buckets in CESR for
feedback or excitation. The iGp has the ability to resolve
individual bunches, however, the system is limited by the electrical
length of the stripline kickers and the field decay time of the
cavity kicker to provide differing feedback kicks 4~nsec apart.

\begin{figure}[htbp] 
   \centering
   \includegraphics[width=0.6\columnwidth]{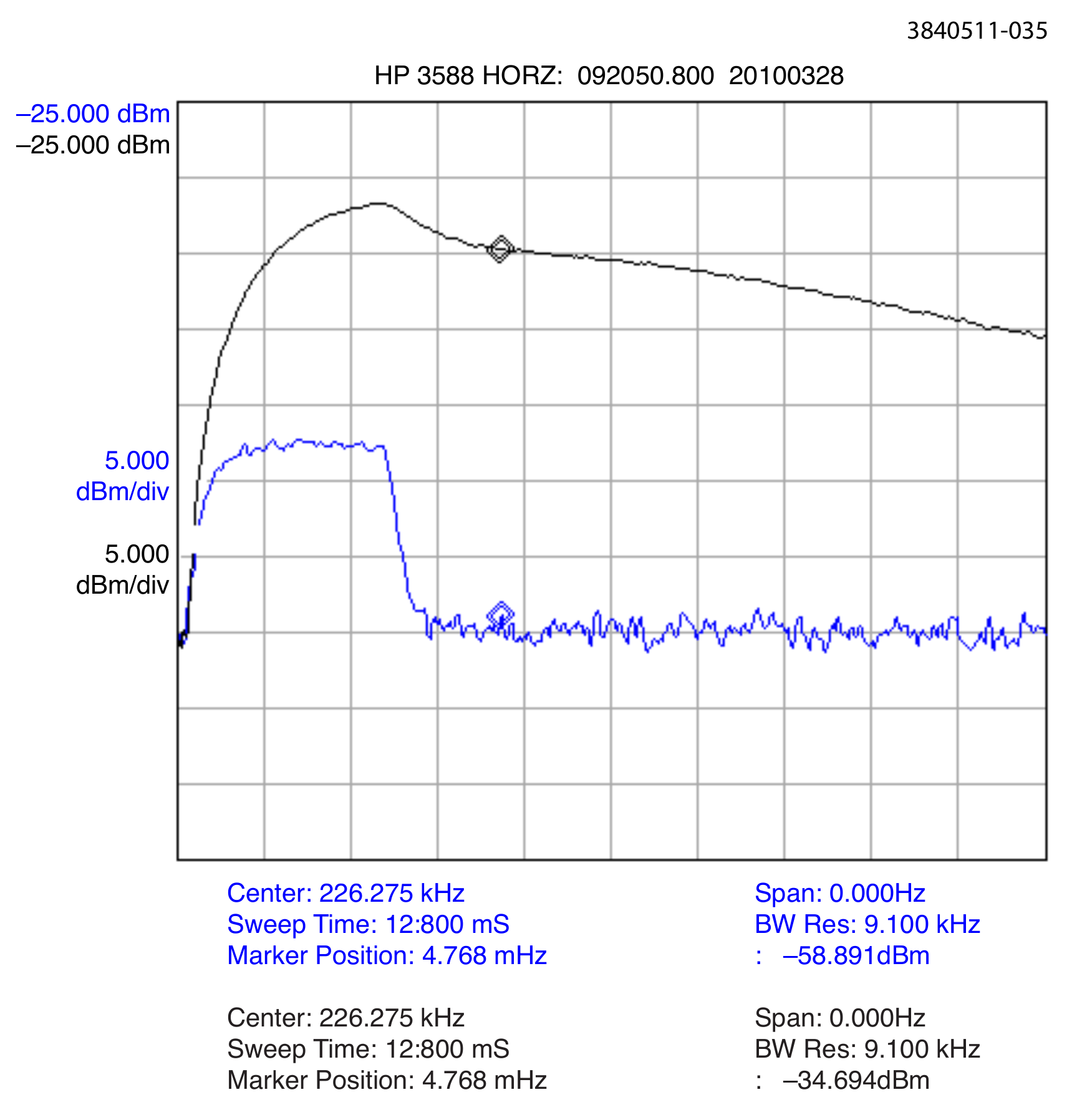}
   \caption[4~nsec horizontal feedback drive-damp measurement]{\label{fig:4nsFbHorzDamp}
   A drive-damp measurement with the horizontal 4~nsec feedback off (black trace) and on (blue trace.)}
\end{figure}

\subsection{Feedback System Results}

The Dimtel iGp upgrade proved successful in stabilizing either
electron or positron beam (separately) in a wide array of bunch
patterns and spacings.  For example, at 5.3~GeV the single bunch
damping in the horizontal system is shown in
Figure~\ref{fig:4nsFbHorzDamp}.  In this drive-damp measurement, the
beam is driven at horizontal betatron frequency for 3~msec and then
allowed to damp.  The black curve shows the natural damping without
any horizontal feedback.  The blue curve shows the result with the
4~nsec feedback on.  The feedback drops the maximum excitation by
about 15~dB and the beam is damped back to its noise floor within
500~$\mu$sec.  The resulting damping times are approximately 9~msec
without feedback and less than 0.32~msec with feedback.

\section{Summary}
\label{sec:summary}

This paper describes the instrumentation, which has been developed
or modified for use in the CesrTA~ program for the investigation of
storage ring beam dynamics.  In particular these studies have
focused on the methods for low emittance tuning of the beam, on the
causes for intra-beam scattering of single bunches and on the
production of ECs, produced by photo-electrons from synchrotron
radiation and secondary emission, and their interaction with bunches
within trains. A companion paper describes how this instrumentation
is used for coordinated beam dynamics measurements.




\bibliographystyle{JHEP}
\bibliography{CesrTA}







\end{document}